\documentclass[floatfix,11pt,nofootinbib]{revtex4}
\usepackage{amssymb,amsmath}
\usepackage{graphicx,float}
\usepackage{indentfirst}

\newcommand{\beq}{\begin{equation}}
\newcommand{\eeq}{\end{equation}}
\newcommand{\bea}{\begin{eqnarray}}
\newcommand{\eea}{\end{eqnarray}}

\def\({\left(}
\def\){\right)}

\input{tcilatex}
\begin{document}

\title{Dynamical heredity from f(R)-bulk to braneworld: curvature dynamical
constraint and f(R)-unimodular gravity}
\author{Andr\'{e} Martorano Kuerten}
\email{martoranokuerten@hotmail.com}
\affiliation{Independent researcher}

\begin{abstract}
Recently, Borzou \textit{et al.} (\textrm{BSSY}) generalized the
Shiromizu-Maeda-Sasaki (\textrm{SMS}) formulation to f(R)-bulks. \textrm{BSSY%
} brane projected equation carries an additional stress tensor, besides 
\textrm{SMS} correction for Einstein's theory on the brane. If we change
this perspective, by requiring \textrm{BSSY}\ tensor in the geometrical side
acting as f(R)-brane generator, it is possible to relate f(R)-brane/bulk
theories, by using curvature dynamical constraint (\textrm{CDC}), a concept
that we developed. Since that brane and bulk are f(R), 5/4D scalar
curvatures also play a dynamical role and, thus, a dynamical version to
Gauss equation trace, or \textrm{CDC}, must be offered. We will work yet in
a specific case to obtain f(R)-unimodular gravity, formally identical with
obtained by Nojiri \textit{et al.} (\textrm{NOO}). Therefore, two
applications which consider cosmological scenarios in f(R)-unimodular
gravity\ with dark radiation correction will be offered.
\end{abstract}

\maketitle

\section{Introduction}

By following pioneer models \cite{rubakov,RS99a,RS99b,gog1,gog2}, \textrm{SMS%
} provided projected Einstein equations onto $3$-thin braneworld \cite%
{sms,sms2} from $5D$ bulk. If we choose \textrm{anti-de Sitter} bulk,
Randall-Sundrum model (\textrm{RS}) with infinite extra dimension \cite%
{RS99b}\ can be obtained by using the structure of \textrm{SMS}. A lot of
these ideas were inspired on string theory advances since \textrm{RS}
scenario can be gotten from Horava-Witten's theory \cite{witten}. \textrm{SMS%
} formulation gives two corrections for usual gravity theory: projected Weyl
tensor $\mathcal{E}_{\mu \nu }$\ and high-energy term $\pi _{\mu \nu }$.
Object $\pi _{\mu \nu }$ must be considered in the early universe when the
quadratic matter-energy density would can overcome the brane tension \cite%
{sms}. Despite $\pi _{\mu \nu }$ changes $4D$ Einstein's theory in the
matter-energy presence, $\mathcal{E}_{\mu \nu }$ modifies vacuum theory, for
instance black holes theories \cite{dadhich,aliev,CFM,neves}. If we consider
cosmological scenarios, both $\pi _{\mu \nu }$ and $\mathcal{E}_{\mu \nu }$
lead to corrections in the Friedmann equation \cite{bruni,govender}, as also
in the black hole metric with electromagnetic radiation \cite{chamblin}.
Actually, $\mathcal{E}_{\mu \nu }$ generates an effective radiation on the
brane (\textit{dark radiation)}, which its source can be, for example, 
\textrm{Schwarzschild}-\textrm{anti-de Sitter} bulk \cite{mukohyama}. This
radiation can generate \textrm{LTB} branes, as shown in \cite{neves}.
Possible connection of $\mathcal{E}_{\mu \nu }$ and $\pi _{\mu \nu }$ with 
\textrm{AdS/CFT} duality can be seen in \cite{shiromizu,papantonopoulos}.

On another hand, dynamical extensions for General Relativity are technicaly
viables. Several possible modifications are given by $f(R)$ theories \cite%
{bergmann,ruzmaikina,breizman,buchdahl,felice,nojiri}. Such theories
substitute Einstein-Hilbert curvature term by a generalized form $f(R)$,
with least action principle yielding alternative field equations. Dynamical
scalar equation controlling an additional freedom degree is derived by
taking in account the trace of these field equations. Sometimes, this
freedom degree is called of scalaron field. We can recognize scalaron with 
\textit{inflaton} field, since our gravity theory is like-Starobinsky: $%
f(R)=R+aR^{2}$ \cite{Starobinsky}. By considering $\mathcal{F(R)}$-bulk and
using \textrm{SMS} procedure, in \cite{borzou}, authors obtained new tensor
components on the brane: $\mathcal{Q}_{\mu \nu }$ (\textrm{BSSY} tensor).
This tensor carries $\mathcal{F(R)}$ functions and together with $\mathcal{E}%
_{\mu \nu }$ and $\pi _{\mu \nu }$ establish the full correction in the
Einstein equations. Cosmological application is given in \cite{borzou}. In
Ref. \cite{larranaga}, it has been analized topological braneworld black
hole with constant scalar curvature. Since trace of $\mathcal{Q}_{\mu \nu }$
appears in the metric of topologically charged black holes as effective
cosmological constant on the brane, solutions have been studied by analising
classical tests of General Relativity in \cite{roldao}.

By wanting to developed some mechanism that makes the projection $\mathcal{%
F(R)}\Rightarrow f(R)$, we will consider \textrm{BSSY} equations, as well as
to announce additionally:%
\begin{equation*}
\text{\textquotedblleft \textit{if }}\mathcal{Q}_{\mu \nu }\text{\textit{\
is in the left side, it generates an effective }}f(R)\text{\textit{-theory
on the brane\textquotedblright .}}
\end{equation*}%
If we consider this assumption, brane-bulk scalaron equation is obtained
naturally, in which $\mathcal{Q=Q}_{\mu }{}^{\mu }$ plays fundamental role.
Since $\mathcal{Q}$ inherits the information of the $\mathcal{F(R)}$
functions, $f(R)$ solution of our mixed scalaron equation is directly
influenced by the $\mathcal{F(R)}$ dynamics. There are two stages to compute:

\textit{First}: $f(R)$-brane can be obtained by expressing some particular
equation to $\mathcal{Q}$, where we do not need the previous knowledge of $%
\mathcal{F(R)}$.

\textit{Second}: if we choose some specific bulk, it implies an extra
equation which reports, dynamically, extrinsic with intrinsic curvature.

Second stage is necessary to find explicitly $\mathcal{F(R)}\Rightarrow f(R)$%
. If taken in account the decomposition $\mathcal{Q}_{\mu \nu }=q_{\mu \nu
}Q $, we will observe yet that \textrm{BSSY} theory implies $f(R)$%
-unimodular gravity, which is formally identical with obtained in \cite%
{noo,noo2}. Nowadays, unimodular gravity \cite{anderson,unruh} gives hope to
solve the cosmological constant problem \cite{weinberg,carroll}. This
problem emerges due to cosmological constant value obtained from General
Relativity or Quantum field theory, since each result does not agree with
each another. Key point is that unimodular gravity generates a cosmological
constant unrelated directly with the usual constant. Strictly speaking, we
will show that our unimodular gravity is obtained by applying traceless
differential operator in $f(R)\neq R$. Taking in account
Friedmann-Robertson-Walker (\textrm{FRW}) ansatz, finally, we will obtain
cosmological expressions.

In the \textrm{section 2} we will review quickly the $f(R)$-unimodular
gravity and \textrm{BSSY} thin brane formulation. Already in the \textrm{%
section 3 }we will promote our main ideas about effective $f(R)$-branes. A
general approach will be elaborate in the \textrm{section 4}, where we will
obtain the $f(R)$-unimodular gravity. In the \textrm{section 5} we will
conclude our study. In order to fix the notation, hereupon, $\{\theta _{\mu
}\}$, with {$\mu =0,1,2,3$} [and $\{\theta _{a}\}$, with {$a=0,1,2,3,5$}]
denotes a basis for the cotangent bundle on a braneworld, embedded in the $%
5D $ bulk. Furthermore, $\{e_{a}\}$ is its dual basis and $\theta
^{a}=dx^{a} $, when a coordinate chart is chosen. Let $n=n_{a}\,\theta ^{a}$
be a timelike covector field normal to the brane and $y$ the associated
Gaussian coordinate. In particular, $n_{a}\,dx^{a}=dy$ on the hypersurface
defined by $y=0$. The brane metric $q_{\mu \nu }$ and the corresponding
components of the bulk metric $g_{ab}$ are in general related by $%
g_{ab}=q_{ab}+n_{a}\,n_{b}$. With these choices it follows that $g_{55}=1$
and $g_{\mu 5}=0$, the $5D$ bulk metric%
\begin{equation}
g_{ab}\,dx^{a}\,dx^{b}=q_{\mu \nu }(x^{\alpha },y)\,dx^{\mu }\,dx^{\nu
}+dy^{2}.
\end{equation}

\section{Short Reviews: $f(R)$-Unimodular Gravity and BSSY Brane Formulation}

\subsection{$f(R)$-Unimodular Gravity}

The combination of $f(R)$ theories with unimodular gravity has been done in 
\cite{noo,noo2}. If we take $4D$ action given by%
\begin{equation}
S=\frac{1}{\kappa _{4}^{2}}\int d^{4}x\left[ \sqrt{-g}\left( f\left(
R\right) -\pounds \right) +\pounds \right] +S_{fields},\text{ \ \ }\kappa
_{4}^{2}\equiv 8\pi G,
\end{equation}%
and so, by varying it with respect to the metric, we obtain%
\begin{equation}
R_{\mu \nu }d_{R}f(R)-\frac{1}{2}\left[ f(R)-\pounds \right] g_{\mu \nu
}+\left( g_{\mu \nu }\square -\nabla _{\mu }\nabla _{\nu }\right)
d_{R}f(R)=\kappa _{4}^{2}T_{\mu \nu },  \label{noo}
\end{equation}%
where $S_{fields}$ and $\pounds $ are, respectively, physical fields action
and Lagrange multipler function (for details \cite{noo}). Still, $g$ is the
metric determinant and $G$ is the usual Newton gravitation constant. We will
use the notation $d_{R}\equiv d/dR$.

When $S$ is varied by $\pounds $, the called unimodular constraint%
\begin{equation}
\sqrt{-g}=1,
\end{equation}%
is derived. In the Refs. \cite{noo,noo2}, have been shown several physical
applications. For instance, in \cite{noo} has been studied inflationary
scenarios while Newton law behavior in \cite{noo,noo2}.

If we want to derive $f(R)$ theories \cite%
{bergmann,ruzmaikina,breizman,buchdahl,felice,nojiri} from (\ref{noo}), we
must choice $\pounds =0$, i. e.,%
\begin{equation}
R_{\mu \nu }d_{R}f(R)-\frac{1}{2}f(R)g_{\mu \nu }+\left( g_{\mu \nu }\square
-\nabla _{\mu }\nabla _{\nu }\right) d_{R}f(R)=\kappa _{4}^{2}T_{\mu \nu },
\label{f(R)o}
\end{equation}%
$f(R)$ theories are scalar-tensor theories, in the sense that trace of (\ref%
{f(R)o}) provides a scalar dynamical equation:%
\begin{equation}
\left[ \left( R+3\square \right) d_{R}-2\right] f(R)=\kappa _{4}^{2}T.
\label{sfe}
\end{equation}%
Expression (\ref{sfe}) has as source the stress tensor trace: $T$. For
example, by putting $f(R)=R+aR^{2}$ in (\ref{sfe}), it becomes%
\begin{equation}
\left( \square -m^{2}\right) R=m^{2}\kappa _{4}^{2}T,\text{ \ \ with \ }%
m\equiv \pm 1/6a^{2}.  \label{stbt}
\end{equation}%
Therefore, curvature scalar $R$ satisfies Klein-Gordon equation with
associated mass $m\equiv 1/6a^{2}$ and source $\kappa _{4}^{2}T$. Therewith,
several authors sometimes have called $R$ (or $d_{R}f(R)$) of scalaron field.

If taken into account Friedmann-Robertson-Walker (\textrm{FRW}) metric and $%
T=0$, expression (\ref{f(R)o}) yields (for details \cite{felice})%
\begin{equation}
\digamma \equiv 6H\partial _{t}\partial _{t}H+18H^{2}\partial _{t}H-3\left(
\partial _{t}H\right) ^{2}=-3m^{2}H^{2},  \label{Re}
\end{equation}%
with $H\equiv a^{-1}\partial _{t}a$ being the Hubble function. Result (\ref%
{Re}) composes Starobinsky theory proposed in 1980 \cite{Starobinsky} and it
is first Friedmann equation in this case. In the inflation epoch, $\digamma 
\mathfrak{\simeq }18H^{2}\partial _{t}H\simeq -3m^{2}H^{2}$, so that $%
H\simeq H_{0}-(m^{2}/6)(t-t_{0})$ leads to an inflationary scale factor $%
a\simeq a_{0}\exp [H_{0}(t-t_{0})-(m^{2}/12)(t-t_{0})^{2}]$ where $H_{0}$
and $a_{0}$ are defined in the start of the inflation $t_{0}$.

Unimodular gravity \cite{anderson,unruh} is obtained by choosing $f(R)=R$ in
(\ref{noo}) and so getting its trace: $2\pounds =\kappa _{4}^{2}T-R$, such
that we can rewrite (\ref{noo}) as follows%
\begin{equation}
R_{\mu \nu }-\frac{1}{4}Rg_{\mu \nu }=\kappa _{4}^{2}\left( T_{\mu \nu }-%
\frac{1}{4}Tg_{\mu \nu }\right) .  \label{ug}
\end{equation}%
Taking the divergence of (\ref{ug}), we have%
\begin{equation}
\nabla ^{\mu }G_{\mu \nu }=0=\nabla ^{\mu }T_{\mu \nu }:\text{ \ \ }\partial
_{\mu }\left( R+\kappa _{4}^{2}T\right) =0,
\end{equation}%
which implies $\Lambda ^{(U)}\equiv R+\kappa _{4}^{2}T$ constant. If now we
substitute $T$ in (\ref{ug}), we obtain then the usual Einstein theory given
by%
\begin{equation}
R_{\mu \nu }-\frac{1}{2}Rg_{\mu \nu }=-\Lambda ^{(U)}g_{\mu \nu }+\kappa
_{4}^{2}T_{\mu \nu }.
\end{equation}%
Since $\Lambda ^{(U)}$ does not related with cosmological constant found in
Einstein theory, it is unrelated to the vacuum energy regarded in Quantum
field theory. Thus, unimodular gravity \textquotedblleft
solves\textquotedblright\ the cosmological constant problem \cite%
{weinberg,carroll}.

\subsection{BSSY Brane Formulation}

By generalizing \textrm{RS} model \cite{RS99a,RS99b}, \textrm{SMS} obtained
effective Einstein equations on thin braneworlds \cite{sms,sms2} (for a
review \cite{maartens}). In a similar way, \textrm{BSSY} equations \cite%
{borzou} are obtained by taking Gauss equations 
\begin{equation}
R_{abc}{}^{f}=q^{f}{}_{e}q{}_{a}{}^{d}q{}_{b}{}^{g}q{}_{c}{}^{h}{}\mathcal{R}%
_{d}{}_{gh}{}^{e}+2K_{c[a}{}{}{}{}K_{b]}{}^{f},  \label{ge}
\end{equation}%
and Israel junction%
\begin{equation}
K_{\mu \nu }=-\frac{1}{2}k_{5}^{2}\left[ \tau _{\mu \nu }+\frac{1}{3}\left(
\lambda -\tau \right) q_{\mu \nu }\right] .  \label{ijc}
\end{equation}%
$\mathcal{R}_{d}{}_{gh}{}^{e}$ is the five dimensional Riemann tensor and $%
R_{abc}{}^{f}$ its four dimensional version. $K_{\mu \nu }$ is the brane
extrinsic curvature in $y=0$ where $K_{\mu \nu }\sim \partial _{y}q_{\mu \nu
}$ at Gaussian coordinates. Equation (\ref{ijc}) is obtained if $\mathbb{Z}%
_{2}$-symmetry is assumed. $5D$ gravitation field equations taken by \textrm{%
BSSY} are%
\begin{equation}
\mathcal{R}_{ab}d_{\mathcal{R}}\mathcal{F}(\mathcal{R})-\frac{1}{2}\mathcal{F%
}(\mathcal{R})g_{ab}+\left( g_{ab}\boxminus -D_{a}D_{b}\right) d_{\mathcal{R}%
}\mathcal{F}(\mathcal{R)}=\kappa _{5}^{2}T_{ab},  \label{5dfr}
\end{equation}%
with $5D$ stress tensor given by%
\begin{equation}
T_{ab}=-\Lambda _{5}g_{ab}+\left( -\lambda g_{ab}+\tau _{ab}\right) \delta
(y).
\end{equation}%
We have used the notation: $d_{\mathcal{R}}\equiv d/d\mathcal{R}$. $D_{a}$
is covariant derivative with respect to bulk metric $g_{ab}$ and $\boxminus
\equiv D^{a}D_{a}$. $\lambda $ and $\tau _{\mu \nu }$ are, respectively,
brane tension and stress tensor of fields on the brane. The function $\delta
(y)$ provides thin brane localization: $y=0$. Original \textrm{SMS}
formulation is derived by taking $\mathcal{F}(\mathcal{R})=\mathcal{R}$ in (%
\ref{5dfr}).

If we combine (\ref{ge}), (\ref{ijc}) and (\ref{5dfr}), these equations
yield the projected equations on the brane \cite{borzou}%
\begin{equation}
R_{\mu \nu }-\frac{1}{2}Rq_{\mu \nu }=\mathcal{J}_{\mu \nu }+\mathcal{Q}%
_{\mu \nu },  \label{fro4D}
\end{equation}%
where%
\begin{equation}
\mathcal{J}_{\mu \nu }=-\Lambda _{4}q_{\mu \nu }+\kappa _{4}^{2}\tau _{\mu
\nu }+\frac{6\kappa _{4}^{2}}{\lambda }\pi _{\mu \nu }-\mathcal{E}_{\mu \nu
}.  \label{sms}
\end{equation}%
$\mathcal{J}_{\mu \nu }$ is the \textrm{SMS} stress tensor with $\pi _{\mu
\nu }$ given by 
\begin{equation}
\pi _{\mu \nu }=\frac{1}{12}\tau \tau _{\mu \nu }-\frac{1}{4}\tau _{\mu
\sigma }\tau _{\nu }^{\sigma }+\frac{1}{24}(3\tau _{\sigma \delta }\tau
^{\sigma \delta }-\tau ^{2})\,q_{\mu \nu },\text{ \ \ }\tau =\tau _{\mu
}{}^{\mu }.
\end{equation}%
Object $\pi _{\mu \nu }$ represents a high energy correction while $\mathcal{%
E}_{\mu \nu }=$ $q_{\mu }{}^{a}q_{\nu }{}^{b}\mathcal{C}%
^{d}{}_{acb}n_{d}n^{c}$ originates from $5D$ Weyl tensor: $\mathcal{C}%
^{d}{}_{acb}$.

Since $\mathcal{E}_{\mu \nu }$ is traceless, it behaves as an effective
radiation on the brane, which carries informations about the bulk geometry 
\cite{sms,maartens,maartens2}. Several physical implications of $\mathcal{E}%
_{\mu \nu }$ are given in the Refs. \cite%
{dadhich,aliev,CFM,neves,bruni,govender,chamblin,mukohyama,shiromizu,papantonopoulos}%
. $\Lambda _{5}$ is the bulk cosmological constant while the effective
cosmological constant $\Lambda _{4}$ is given by%
\begin{equation}
\Lambda _{4}=\frac{1}{2}k_{5}^{2}\left[ \Lambda _{5}+\frac{1}{6}%
k_{5}^{2}\lambda ^{2}\right] ,\text{ \ \ with }\kappa _{4}^{2}=\frac{\lambda 
}{6}\kappa _{5}^{4},  \label{4Dcc}
\end{equation}%
where $\kappa _{4}$ ($\kappa _{5}$) is the four (five) dimensional
gravitation constant. Randall-Sundrum model \cite{RS99b}\ is obtained by
taking in account $\Lambda _{4}=0$, since that $\Lambda _{5}<0$ from (\ref%
{4Dcc}).

If we put $\mathcal{F}(\mathcal{R})=\mathcal{R}$ in (\ref{5dfr}), we have $%
\mathcal{Q}_{\mu \nu }=0$ in (\ref{fro4D}), such that the term $\mathcal{Q}%
_{\mu \nu }$ acts\ as a stress tensor on the brane providing new corrections
besides \textrm{SMS}. Consequently, we have three corrections in the
Einstein equations for $\mathcal{F}(\mathcal{R})\neq \mathcal{R}$ case.
Explicitly, $\mathcal{Q}_{\mu \nu }$ is given by%
\begin{equation}
\mathcal{Q}_{\mu \nu }=\left[ F(\mathcal{R})q_{\mu \nu }+\frac{2}{3}\frac{%
D_{a}D_{b}\left( d_{\mathcal{R}}\mathcal{F}\right) }{d_{\mathcal{R}}\mathcal{%
F}}\left( \delta _{\mu }^{a}\delta _{\nu }^{b}+n^{a}n^{b}q_{\mu \nu }\right) %
\right] _{y=0},
\end{equation}%
with $F(\mathcal{R})$ being%
\begin{equation}
F(\mathcal{R})=-\frac{4}{15}\frac{\boxminus \left( d_{\mathcal{R}}\mathcal{F}%
\right) }{d_{\mathcal{R}}\mathcal{F}}-\frac{\mathcal{R}}{10}\left( \frac{3}{2%
}+d_{\mathcal{R}}\mathcal{F}\right) +\frac{1}{4}\mathcal{F-}\frac{2}{5}%
\boxminus \left( d_{\mathcal{R}}\mathcal{F}\right) .
\end{equation}%
We should note that $\mathcal{Q}_{\mu \nu }$ encapsulates the $\mathcal{F(R)}
$ effects on the brane.

By taking the trace of (\ref{fro4D}), we have%
\begin{equation}
R=-\mathcal{Q-J},  \label{ft}
\end{equation}%
with%
\begin{equation}
\mathcal{J}=-4\Lambda _{4}+\kappa _{4}^{2}\tau +\frac{6\kappa _{4}^{2}}{%
\lambda }\pi _{\mu }{}^{\mu }\text{ \ \ and \ }\mathcal{Q}=\left[ 4F(%
\mathcal{R})+\frac{2}{3}\frac{D_{a}D_{b}\left( d_{\mathcal{R}}\mathcal{F}%
\right) }{d_{\mathcal{R}}\mathcal{F}}\left( \delta ^{ab}+4n^{a}n^{b}\right) %
\right] _{y=0}.  \label{qt}
\end{equation}%
Null divergence requeriment becomes%
\begin{equation}
\nabla ^{\mu }\mathcal{Q}_{\mu \nu }=\nabla ^{\mu }\left( \mathcal{E}_{\mu
\nu }-\frac{6\kappa _{4}^{2}}{\lambda }\pi _{\mu \nu }\right) ,  \label{nd}
\end{equation}%
since that $\nabla ^{\mu }\left( \mathcal{J}_{\mu \nu }+\mathcal{Q}_{\mu \nu
}\right) =0$. Onto a vacuum conformally flat bulk, we have $\nabla ^{\mu }%
\mathcal{Q}_{\mu \nu }=0$, so that we can identify $\mathcal{Q}_{\mu \nu }$
as a kind of matter. Several approaches projecting $f(R)$-bulk in thin
braneworlds can be seen in the Refs. \cite{carames,haghani,Chakraborty}.

\section{$f(R)$-branes and Curvature Dynamical Constraint}

\subsection{$f(R)$-branes}

We have seen that $\mathcal{F(R)}$-bulk yields an additional extra stress
tensor in Einstein brane. Our proposal is to create a mechanism in which $%
\mathcal{F(R)}$-bulk generates an effective $f(R)$-brane.

Instead to consider the original idea (\ref{fro4D}), we will require that $%
\mathcal{Q}_{\mu \nu }$ operates on the left side of (\ref{fro4D}) in the
following way%
\begin{equation}
G_{\mu \nu }^{[f(R)]}\equiv G_{\mu \nu }^{[R]}-\mathcal{Q}_{\mu \nu }=%
\mathcal{J}_{\mu \nu }.  \label{aa}
\end{equation}%
Notation $G_{\mu \nu }^{[f(R)]}$ denotes Einstein tensor component for some
four dimensional $f(R)$ theory. Therefore, we can rewrite (\ref{fro4D}) and (%
\ref{aa}) as 
\begin{equation}
R_{\mu \nu }-\frac{1}{2}Rq_{\mu \nu }-\mathcal{Q}_{\mu \nu }=R_{\mu \nu
}d_{R}f-\frac{1}{2}fq_{\mu \nu }+\left( q_{\mu \nu }\square -\nabla _{\mu
}\nabla _{\nu }\right) d_{R}f,  \label{ab}
\end{equation}%
\begin{equation}
R_{\mu \nu }d_{R}f-\frac{1}{2}fq_{\mu \nu }+\left( q_{\mu \nu }\square
-\nabla _{\mu }\nabla _{\nu }\right) d_{R}f=\mathcal{J}_{\mu \nu },
\label{ac}
\end{equation}%
with each respective trace, respectively, yielding%
\begin{equation}
\left[ \left( R+3\square \right) d_{R}-2\right] f(R)+R=-\mathcal{Q},
\label{ke}
\end{equation}%
\begin{equation}
\left[ \left( R+3\square \right) d_{R}-2\right] f(R)=\mathcal{J}.
\label{ke2}
\end{equation}%
Expression (\ref{ke}) can be obtained directly of (\ref{ke2}), by using (\ref%
{ft}). Equation (\ref{ke2}) establishes a \textrm{SMS} version to scalaron
dynamical equation (\ref{sfe}). Explicitly in the vacuum, we have not any
difference with scalaron theory without extra dimension. This result gives
the information which scalaron field does not explicitly affected by $%
\mathcal{E}_{\mu \nu }$. If $\tau \neq 0$, we must consider $\pi _{\alpha
}{}^{\alpha }$.

If we want to obtain an effective $f(R)$-brane, let us consider (\ref{ke}).
Since that%
\begin{equation}
\mathcal{Q}=\mathcal{Q}(\mathcal{F(}\mathcal{R(}R\mathcal{))}),
\end{equation}%
with $\mathcal{R}(R)$ dictated by the Gauss equations trace (\ref{ge}):%
\begin{equation}
\mathcal{R}{}={}R+\mathcal{K},\text{ \ \ with \ \ }\mathcal{K}%
=K_{ab}{}{}{}{}K^{ab}-K{}{}{}{}^{2},  \label{get}
\end{equation}%
equation (\ref{ke}) can be rewritten as%
\begin{equation}
\Pi ^{\left[ f(R)\right] }f(R)+R=\left[ \frac{2}{5}\Pi ^{\left[ \mathcal{F}%
\left( \mathcal{R}\right) \right] }\mathcal{F}\left( \mathcal{R}\right) +%
\frac{3\mathcal{R}}{5}-\left( d_{\mathcal{R}}\mathcal{F}\right) ^{-1}%
\mathcal{O}\left( d_{\mathcal{R}}\mathcal{F}\right) \right] _{y=0},
\label{54st}
\end{equation}%
where we have used (\ref{qt}). $\mathcal{O}$, $\Pi ^{\left[ f(R)\right] }$
and $\Pi ^{\left[ \mathcal{F}\left( \mathcal{R}\right) \right] }$ are
respectively%
\begin{equation}
\mathcal{O}\equiv \frac{2}{3}\left( \delta ^{ab}+4n^{a}n^{b}\right)
D_{a}D_{b}-\frac{16}{15}\boxminus ,  \label{op1}
\end{equation}%
\begin{equation}
\Pi ^{\left[ f(R)\right] }\equiv \left( R+3\square \right) d_{R}-2\text{ \ \
and \ }\ \Pi ^{\left[ \mathcal{F}\left( \mathcal{R}\right) \right] }\equiv
\left( \mathcal{R}+4\boxminus \right) d\mathcal{_{R}}-\frac{5}{2}.
\end{equation}%
If we take the trace of (\ref{5dfr}), we obtain the $5D$ scalaron theory%
\begin{equation}
\Pi ^{\left[ \mathcal{F}\left( \mathcal{R}\right) \right] }\mathcal{F}\left( 
\mathcal{R}\right) =\kappa _{5}^{2}T,  \label{5ds}
\end{equation}%
so that $\Pi ^{\left[ f(R)\right] }$ (by (\ref{ke2})) and $\Pi ^{\left[ 
\mathcal{F}\left( \mathcal{R}\right) \right] }$ are, respectively, brane
(bulk) scalaron operators. Thus, expression (\ref{54st}) provides a relation
between $5D$ scalaron theory with its $4D$ version. It is easy to see that
if we consider $\mathcal{F}(\mathcal{R})=\mathcal{R}$ ($\mathcal{Q}=0$),
then $f(R)=R$. This result gives the original \textrm{SMS} theory.

Now, let us work non-trivial examples. If the equation $3\square d_{R}f(R)=-%
\mathcal{Q}$ is satisfied in (\ref{ke}), we must solve $\left(
-Rd_{R}+2\right) f=1$, which implies a \textquotedblleft
Starobinsky-Shiromizu-Maeda-Sazaki\textquotedblright\ brane (\textrm{SSMS}
brane). In fact%
\begin{equation}
3\square d_{R}f(R)=-\mathcal{Q}:\text{ \ \ }Rd_{R}f(R)-2f(R)+1=0\text{ \ \ }%
\mapsto \text{ \ \ }f(R)=R+\mathfrak{a}R^{2},  \label{gb}
\end{equation}%
with $\mathfrak{a}$ an arbitrary constant. Therefore, \textrm{SSMS} brane
can be represented by the equations%
\begin{equation}
G_{\mu \nu }^{[R+\mathfrak{a}R^{2}]}=\mathcal{J}_{\mu \nu }=-\Lambda
_{4}q_{\mu \nu }+\kappa _{4}^{2}\tau _{\mu \nu }+\frac{6\kappa _{4}^{2}}{%
\lambda }\pi _{\mu \nu }-\mathcal{E}_{\mu \nu },  \label{ssms}
\end{equation}%
\begin{equation}
\mathfrak{a}\square R=-\frac{1}{6}\left[ \frac{2}{5}\Pi ^{\left[ \mathcal{F}%
\left( \mathcal{R}\right) \right] }\mathcal{F}\left( \mathcal{R}\right) +%
\frac{3\mathcal{R}}{5}-\left( d_{\mathcal{R}}\mathcal{F}\right) ^{-1}%
\mathcal{O}\left( d_{\mathcal{R}}\mathcal{F}\right) \right] _{y=0}.
\label{ssms2}
\end{equation}%
Thus, hybrid theory which combines inflationary Starobinsky model with 
\textrm{SMS} theory is obtained from (\ref{ssms}) and (\ref{ssms2}). If $%
\mathfrak{a}=0$, we recover \textrm{SMS} formulation. By using (\ref{ssms})
and (\ref{ssms2}), we obtain alternative forms to \textrm{SSMS }scalaron
theory, i.e.,%
\begin{equation}
\Delta _{KG}^{\left( -\right) }R=m^{2}\mathcal{J}\text{ \ \ or \ \ }\square
R=-m^{2}\mathcal{Q}\text{ \ \ or \ }\Delta _{KG}^{\left( -\right) }\mathcal{Q%
}=-\square \mathcal{J},  \label{st1}
\end{equation}%
with $m^{2}\equiv 1/6\mathfrak{a}$. $\Delta _{KG}^{\left( \pm \right)
}\equiv \square \pm m^{2}$ is the positive (negative) Klein Gordon operator.

Others two examples are given by%
\begin{equation}
3\square d_{R}f+R=-\mathcal{Q}:\text{\ }Rd_{R}f=2f\text{ \ }\mapsto \text{ \ 
}f(R)=\mathfrak{f}R^{2}\text{ \ }\mapsto \text{\ \ }\Delta _{KG}^{\left(
+\right) }R=-m^{2}\mathcal{Q},
\end{equation}%
\begin{equation}
3\square d_{R}f-2f=-\mathcal{Q}:\text{\ }d_{R}f=-1\text{ \ }\mapsto \text{ \ 
}f(R)=-R+\mathfrak{c}\text{ \ }\mapsto \text{\ \ }2R+\mathcal{Q}=2\mathfrak{c%
},  \label{esms}
\end{equation}%
where $\mathfrak{f}=f_{0}/R_{0}^{2}$ and $\mathfrak{c}$\ are constants. In
this stage, we do not work bulk theory. In the follows, we will developed a
mecanism which relates brane with bulk theories.

\subsection{Projecting $\mathcal{F}(\mathcal{R})\Rightarrow f(R)$: Curvature
Dynamical Constraint}

By introducing the concept of curvature dynamical constraint (\textrm{CDC}),
we will develop a mechanism which provides an effective $f(R)$-brane from $%
\mathcal{F(R)}$-bulk. Equation (\ref{get}) will be called of curvature
geometrical constraint (\textrm{CGC}). \textrm{CDC} is necessary if we want
to know explicitly the $\mathcal{F(R)}$-bulk. The philosophical idea is that
if we require any gravitation theory with specific dynamics on the brane, we
must offer also a \textrm{CDC} besides \textrm{CGC}. For example, if we have
a bulk with $\mathcal{F(R)}$ dynamics which projects a $f(R)$ dynamics on
the brane, it is necessary to provide an extra relation between the
curvature objects.

It is convenient to define the objects $\Pi _{\ast }^{\left[ f\right] }$, $%
\Pi _{\ast }^{\left[ \mathcal{F}\right] }$ and $\Theta $\ as follow%
\begin{equation}
\Pi _{\ast }^{\left[ f(R)\right] }\equiv \Pi ^{\left[ f(R)\right] }+\frac{R}{%
f},\text{ \ \ }\Pi _{\ast }^{\left[ \mathcal{F}\left( \mathcal{R}\right) %
\right] }\equiv \frac{2}{5}\Pi ^{\left[ \mathcal{F}\left( \mathcal{R}\right) %
\right] }+\frac{3\mathcal{R}}{5\mathcal{F}}\text{ \ \ and \ }\Theta \equiv 
\frac{\mathcal{O}d_{\mathcal{R}}\mathcal{F}}{d_{\mathcal{R}}\mathcal{F}}.
\label{def}
\end{equation}%
With (\ref{def}), the triad of equations (\ref{ke}), (\ref{ke2}) and (\ref%
{5ds}) assume the compact forms%
\begin{equation}
\Pi ^{\left[ f(R)\right] }f(R)=\mathcal{J},\text{ \ \ }\Pi _{\ast }^{\left[
f(R)\right] }f(R)=-\mathcal{Q}\text{ \ \ and \ \ }\Pi ^{\left[ \mathcal{F}%
\left( \mathcal{R}\right) \right] }\mathcal{F}\left( \mathcal{R}\right)
=\kappa _{5}^{2}T,  \label{tri}
\end{equation}%
while (\ref{54st}) is rewritten as%
\begin{equation}
\Pi _{\ast }^{\left[ f(R)\right] }f(R)=\left[ \Pi _{\ast }^{\left[ \mathcal{F%
}\left( \mathcal{R}\right) \right] }\mathcal{F}\left( \mathcal{R}\right)
-\Theta \left( \mathcal{R}\right) \right] _{y=0}.  \label{fse}
\end{equation}%
We are now ready to introduce the \textrm{CDC} concept.

Formally, \textrm{CDC} is obtained if the bulk scalaron term $\Pi _{\ast }^{%
\left[ \mathcal{F}\right] }\mathcal{F}$ taken in $y=0$, follows the
prescription 
\begin{equation}
\left[ \Pi _{\ast }^{\left[ \mathcal{F}\left( \mathcal{R}\right) \right] }%
\mathcal{F}\left( \mathcal{R}\right) \right] _{y=0}=\Pi _{\ast }^{\left[ f(R)%
\right] }f(R)+\left[ D\left( R,\mathcal{K}\right) \right] _{y=0}.
\label{fse2}
\end{equation}%
If we put (\ref{fse2}) in (\ref{fse}), we obtain%
\begin{equation}
D\left( R,\mathcal{K}\right) =\Theta \left( \mathcal{R}\right) .  \label{cdc}
\end{equation}%
Object $D\left( \mathcal{K},R\right) $ is some dynamical term relating the
curvatute scalars. Therefore, (\ref{cdc}) is our generic \textrm{CDC}.
Equation (\ref{cdc}) relates only two dynamical variables, since \textrm{CGC}
eliminates one of them.

Let us elaborate an example which considers a $\left( \mathcal{R}+\mathfrak{b%
}\mathcal{R}^{2}\right) $-bulk projecting a $\left( R+\mathfrak{a}%
R^{2}\right) $-brane. By acting $\Pi _{\ast }$-operators on their respective
theories, we have%
\begin{equation}
\Pi _{\ast }^{\left[ R+\mathfrak{a}R^{2}\right] }f=6\mathfrak{a}\square R%
\text{ \ \ and \ \ }\Pi _{\ast }^{\left[ \mathcal{R}+\mathfrak{b}\mathcal{R}%
^{2}\right] }\mathcal{F}=\frac{1}{5}\left[ 16\mathfrak{b}\boxminus -%
\mathfrak{b}\mathcal{R}\right] \mathcal{R}.  \label{fb2}
\end{equation}%
If we take into account (\ref{fb2}), (\ref{get}) and so put them in (\ref%
{fse}), we obtain the following equation%
\begin{equation}
6\mathfrak{a}\square R=\frac{16\mathfrak{b}}{5}\square R+\left[ \frac{%
\mathfrak{b}}{5}\left( 16\boxminus \mathcal{K+}16D^{y}D_{y}R-\mathcal{R}%
^{2}\right) -\frac{2\mathfrak{b}}{\left( 1+2\mathfrak{b}\mathcal{R}\right) }%
\mathcal{OR}\right] _{y=0}.  \label{fsee}
\end{equation}%
We have used $\boxminus =\square +D^{y}D_{y}$ and%
\begin{equation}
\Theta \left( \mathcal{R}\right) =\frac{2\mathfrak{b}}{\left( 1+2\mathfrak{b}%
\mathcal{R}\right) }\mathcal{OR}.
\end{equation}%
By comparing left/right sides of (\ref{fsee}), we note the relation $%
\mathfrak{a}=8\mathfrak{b}/15$. Since that $[\ldots ]_{y=0}$ term is equal
to zero, we find explicitly our first \textrm{CDC}:%
\begin{equation}
\boxminus \mathcal{K+}D^{y}D_{y}R=\left[ \frac{5}{8\left( 1+2\mathfrak{b}%
\mathcal{R}\right) }\mathcal{O+}\frac{1}{16}\mathcal{R}\right] \mathcal{R}.
\label{cdc1}
\end{equation}%
Our result can be formulate as follows: the projection%
\begin{equation*}
\mathcal{F}\left( \mathcal{R}\right) =\mathcal{R}+\mathfrak{b}\mathcal{R}^{2}%
\text{ \ \ }\Rightarrow \text{ \ \ }f(R)=R+(8\mathfrak{b}/15)R^{2},
\end{equation*}%
which has the bulk/brane scalaron theories%
\begin{equation}
\left( \boxminus -m_{5}^{2}-\frac{1}{16}\mathcal{R}\right) \mathcal{R}=\frac{%
2}{3}\kappa _{5}^{2}m_{5}^{2}T\text{ \ \ \ and \ \ }\Delta _{KG}^{\left(
-\right) }R=m_{4}^{2}\mathcal{J}\text{\ },  \label{n4}
\end{equation}%
contains the additional relation (\ref{cdc1}). It is viable to note that $%
5D/4D$ masses, associated with $5D/4D$ scalaron fields, are related by $%
m_{5}^{2}=(3/5)m_{4}^{2}=3/16\mathfrak{b}$.

Let us consider other possible projections which involve $\mathfrak{f}R^{2}$/%
$\mathfrak{h}\mathcal{R}^{2}$-theories. Their $\Pi _{\ast }$-operators
acting on $\mathfrak{f}R^{2}$/$\mathfrak{h}\mathcal{R}^{2}$ are 
\begin{equation}
\Pi _{\ast }^{\left[ \mathfrak{f}R^{2}\right] }f=\left( 6\mathfrak{f}\square
+1\right) R\text{ \ \ and \ \ }\Pi _{\ast }^{\left[ \mathfrak{h}\mathcal{R}%
^{2}\right] }\mathcal{F}=\frac{1}{5}\left[ 16\mathfrak{h}\boxminus +2-%
\mathfrak{h}\mathcal{R}\right] \mathcal{R}.  \label{fb}
\end{equation}%
If we take (\ref{fb}) and (\ref{fb2}), and to consider their full
combinations, posteriorly putting them in (\ref{fse}), we obtain separately
each projection with its respective \textrm{CDC}. In resume, we have%
\begin{equation*}
\begin{tabular}{|l|l|}
\hline
$\mathcal{F}\left( \mathcal{R}\right) \text{ \ \ }\Rightarrow \text{ \ \ }%
f(R)$ & Curvature Dynamical Constraint \\ \hline
$\mathfrak{h}\mathcal{R}^{2}\Rightarrow \ \frac{8\mathfrak{h}}{15}R^{2}$ & $%
\left( \boxminus -\frac{1}{2\mathfrak{h}}\right) \mathcal{K}+\left(
D^{y}D_{y}-\frac{3}{16\mathfrak{h}}\right) R=\frac{1}{16}\left( \mathcal{R}+%
\frac{5}{\mathfrak{h}}\frac{\mathcal{O}}{\mathcal{R}}\right) \mathcal{R}$ \\ 
\hline
$\mathfrak{h}\mathcal{R}^{2}\Rightarrow R+\frac{8\mathfrak{h}}{15}R^{2}$ & $%
\left( \boxminus +\frac{1}{8\mathfrak{h}}\right) \mathcal{K}+\left(
D^{y}D_{y}+\frac{1}{8\mathfrak{h}}\right) R=\frac{1}{16}\left( \mathcal{R}+%
\frac{5}{\mathfrak{h}}\frac{\mathcal{O}}{\mathcal{R}}\right) \mathcal{R}$ \\ 
\hline
$\mathcal{R}+\mathfrak{b}\mathcal{R}^{2}\Rightarrow \frac{8\mathfrak{b}}{15}%
R^{2}$ & $\left( \boxminus -\frac{5}{8\mathfrak{b}}\right) \mathcal{K}%
+\left( D^{y}D_{y}-\frac{5}{16\mathfrak{b}}\right) R=\frac{1}{16}\left( 
\mathcal{R}+\frac{10\mathcal{O}}{1+2\mathfrak{b}\mathcal{R}}\right) \mathcal{%
R}$ \\ \hline
$\mathcal{R}+\mathfrak{b}\mathcal{R}^{2}\Rightarrow R+\frac{8\mathfrak{b}}{15%
}R^{2}$ & $\boxminus \mathcal{K+}D^{y}D_{y}R=\frac{1}{16}\left( \mathcal{R}+%
\frac{10\mathcal{O}}{1+2\mathfrak{b}\mathcal{R}}\right) \mathcal{R}$ \\ 
\hline
\end{tabular}%
\end{equation*}%
In what follows, we will obtain a generalized unimodular if taken into
account a specific case.

\section{Emergent $f(R)$-Unimodular Gravity: $\mathcal{Q}_{\protect\mu 
\protect\nu }=Q(\mathcal{R})q_{\protect\mu \protect\nu }$ case}

In the present section, we will explore the case: $\mathcal{Q}_{\mu \nu }=Q(%
\mathcal{R})q_{\mu \nu }$. This decomposition is possible if%
\begin{equation}
\left[ \frac{2}{3}\delta _{\mu }^{a}\delta _{\nu }^{b}D_{a}D_{b}+q_{\mu \nu
}\left( \frac{2}{3}n^{a}n^{b}D_{a}D_{b}-\frac{4}{15}\boxminus \right) \right]
\left( d_{\mathcal{R}}\mathcal{F}\right) =q_{\mu \nu }\mathcal{O}_{\ast
}\left( d_{\mathcal{R}}\mathcal{F}\right) .
\end{equation}%
In this specific case, we can rewrite (\ref{fro4D}) and (\ref{tri}),
respectively, as 
\begin{equation}
R_{\mu \nu }-\frac{1}{2}Rq_{\mu \nu }=\mathcal{J}_{\mu \nu }+q_{\mu \nu }Q%
\text{ \ \ and \ \ }\Pi _{\ast }^{\left[ f(R)\right] }f(R)=-4Q.  \label{qQ}
\end{equation}%
with $Q$\ given by%
\begin{equation}
Q=\left[ \Theta _{\ast }\left( \mathcal{R}\right) -\frac{1}{4}\Pi _{\ast }^{%
\left[ \mathcal{F}\left( \mathcal{R}\right) \right] }\mathcal{F}\left( 
\mathcal{R}\right) \right] _{y=0}\text{ },\text{ \ where \ }\Theta _{\ast
}\equiv \left( d_{\mathcal{R}}\mathcal{F}\right) ^{-1}\mathcal{O}_{\ast
}\left( d_{\mathcal{R}}\mathcal{F}\right) .
\end{equation}%
Before we obtain $f(R)$-unimodular gravity, let us understand the traceless
Einstein tensor basics for the case (\ref{qQ}).

We will denote by the symbol $^{\circ }$, the composition:%
\begin{equation}
A_{\mu \nu }^{\circ }\equiv A_{\mu \nu }-\frac{1}{4}Aq_{\mu \nu }.
\end{equation}%
for any generic tensor $A_{\mu \nu }$. We should note that $A_{\mu \nu
}^{\circ }$ is traceless. By (\ref{qQ}), we have the traceless Einstein
tensor:%
\begin{equation}
R_{\mu \nu }^{\circ }=\mathcal{J}_{\mu \nu }^{\circ }.  \label{S}
\end{equation}%
Expression (\ref{S}) is obtained directly from tensor equation given in (\ref%
{qQ}). It is easy verify that (\ref{S}) does not contain $\Lambda _{4}$
predicted in \textrm{SMS} formulation, since $q_{\mu \nu }^{\circ }=0$.
Therefore, we have two corrections in the traceless stress-tensor: $\pi
_{\mu \nu }^{\circ }$ and $\mathcal{E}_{\mu \nu }^{\circ }$, which are%
\begin{equation}
\pi _{\mu \nu }^{\circ }=\frac{1}{4}\left[ \frac{1}{3}\tau \tau _{\mu \nu
}-\tau _{\mu \sigma }\tau _{\nu }^{\sigma }+\frac{1}{12}\left( 3\tau
_{\sigma \delta }\tau ^{\sigma \delta }-\tau ^{2}\right) \,q_{\mu \nu }%
\right] \text{ \ and \ }\mathcal{E}_{\mu \nu }^{\circ }=\mathcal{E}_{\mu \nu
}.
\end{equation}%
In the vacuum, equation (\ref{S}) is driven by $\mathcal{E}_{\mu \nu }$ 
\footnote{%
In the usual case: $R_{\mu \nu }^{\circ }=0$.}, i.e.,%
\begin{equation}
R_{\mu \nu }^{\circ }=-\mathcal{E}_{\mu \nu }.  \label{ttv}
\end{equation}%
If we want to obtain $f(R)$-unimodular gravity on the brane, we must combine
the equations (\ref{qQ}) and, posteriorly, to put $\mathcal{J}_{\mu \nu }$
in (\ref{ac}).

By performing the mentioned procedures, we derive generalized unimodular
gravity: 
\begin{equation}
\Delta _{\mu \nu }^{\circ }\left( d_{R}f\right) =R_{\mu \nu }^{\circ }\left(
d_{R}f-1\right) ,  \label{fe}
\end{equation}%
with $\Delta _{\mu \nu }\equiv \nabla _{\mu }\nabla _{\nu }$, so that $%
\Delta _{\mu \nu }^{\circ }$ is our $4D$ traceless operator. Theory (\ref{fe}%
) is supplemented by the equations (\ref{S}). Mathematically, $f(R)$%
-unimodular gravity is stated by the pair: 
\begin{equation}
\left[ R_{\mu \nu }^{\circ }-\Delta _{\mu \nu }^{\circ }\right]
d_{R}f(R)=R_{\mu \nu }^{\circ }\text{ \ \ and \ \ }R_{\mu \nu }^{\circ }=%
\mathcal{J}_{\mu \nu }^{\circ }.  \label{fem}
\end{equation}%
If $f(R)\rightarrow R$, left expression assumes right equation form.

Equations (\ref{fem}) are formally identical with (\ref{noo}). In fact, by
taking the trace of (\ref{noo}) to isolate $\pounds $ and, putting it in (%
\ref{noo}), we obtain (\ref{fem}) with $\kappa _{4}^{2}T_{\mu \nu }^{\circ }$
instead of $\mathcal{J}_{\mu \nu }^{\circ }$. Consequently, our result
provides corrections for \textrm{NOO }theory. For instance, let us consider $%
\Delta T_{\mu \nu }$ a deviation of (\ref{noo}). Since we have now $T_{\mu
\nu }+\Delta T_{\mu \nu }$, we find 
\begin{equation}
\left( R_{\mu \nu }^{\circ }-\Delta _{\mu \nu }^{\circ }\right)
d_{R}f(R)=\kappa _{4}^{2}T_{\mu \nu }^{\circ }+\Delta T_{\mu \nu }^{\circ }.
\label{noo3}
\end{equation}%
If we compare (\ref{noo3}) with (\ref{fem}), we can identify $\Delta T_{\mu
\nu }^{\circ }$ with $\pi _{\mu \nu }^{\circ }$ and $\mathcal{E}_{\mu \nu }$
or, in general,%
\begin{equation}
T_{\mu \nu }^{\circ }\sim \tau _{\mu \nu }^{\circ }\text{ \ \ and \ \ }%
\Delta T_{\mu \nu }^{\circ }\sim \frac{6\kappa _{4}^{2}}{\lambda }\pi _{\mu
\nu }^{\circ }-\mathcal{E}_{\mu \nu }.  \label{noo4}
\end{equation}%
Therefore, $\Delta T_{\mu \nu }^{\circ }$ provides extra dimension signature
if compared with \textrm{NOO} theory. On the vacuum, any deviation can be
speculated as dark radiation manifestation, since $\Delta T_{\mu \nu
}^{\circ }\sim -\mathcal{E}_{\mu \nu }$.

Similarly as has been done in the subsection \textrm{II.A}, we obtain from (%
\ref{S}) the continuity relations%
\begin{equation}
\nabla ^{\mu }\mathcal{J}_{\mu \nu }=\frac{1}{4}\partial _{\nu }\left( 
\mathcal{J}+R\right) =-\partial _{\nu }Q,  \label{nd2}
\end{equation}%
in accordance with (\ref{ft}) and (\ref{nd}). If $\partial _{\mu }Q=0$, we
have $\mathcal{J}+R=-4Q$ constant: $c\equiv 4\Lambda ^{\left( Q\right) }$.
By isoling $\mathcal{J}$ and, posteriorly, by substituting it in (\ref{S}),
we obtain%
\begin{equation}
G_{\mu \nu }^{\left[ R\right] }=-\Lambda ^{\left( Q\right) }q_{\mu \nu }+%
\mathcal{J}_{\mu \nu },\text{ \ \ with \ \ }\Lambda ^{\left( Q\right) }\sim
\Lambda ^{(U)}.  \label{uni}
\end{equation}%
Thus, being $Q$ constant, we can identify it as the cosmological constant
obtained from unimodular theory, while the effective cosmological constant
is $\Lambda _{4}^{eff}\equiv \Lambda _{4}+\Lambda ^{\left( Q\right) }$.

Let us define $\mathfrak{A}(R)$ as follows%
\begin{equation}
\mathfrak{A}(R)\equiv d_{R}f(R)-1,  \label{adef}
\end{equation}%
so that $\mathfrak{A}$ excludes $f(R)=R$ model on the brane. Thus,
expression (\ref{fem}) can be rewritten as%
\begin{equation}
\Delta _{\mu \nu }^{\circ }\mathfrak{A}=R_{\mu \nu }^{\circ }\mathfrak{A}%
\text{ \ \ or \ \ }R_{\mu \nu }^{\circ }=\mathcal{J}_{\mu \nu }^{\circ }=%
\frac{1}{\mathfrak{A}}\Delta _{\mu \nu }^{\circ }\mathfrak{A},\text{ \ \ for
\ \ }\mathfrak{A}\neq 0.  \label{pot}
\end{equation}

It is useful to define also the objects $\Psi _{\mu \nu }$ and $\Phi $:%
\begin{equation}
\Psi _{\mu \nu }\equiv \frac{1}{\mathfrak{A}}\Delta _{\mu \nu }\mathfrak{A}%
\text{ \ \ and \ \ }\Phi \equiv \frac{1}{3}\left[ \frac{2f}{\mathfrak{A}}-R+%
\frac{4Q}{\mathfrak{A}}\right] .  \label{abc}
\end{equation}%
By rewriting (\ref{ke}) with (\ref{adef}) and, simultaneously, taking in
account $\Psi _{\mu }{}^{\mu }$, we obtain%
\begin{equation}
\square \mathfrak{A}=\left( \Psi _{\mu }{}^{\mu }\right) \mathfrak{A}=\Phi 
\mathfrak{A}\text{ \ \ }\mapsto \text{ \ \ }\Psi =\Phi ,
\end{equation}%
such that our brane theory has been codificade in $\Psi _{\mu \nu }$.
Therefore, if we know the information contained in $\Psi _{\mu \nu }$, the
pair $\left[ q^{\mu \nu },\circ \right] $ generates the equations of our $%
f(R)$-unimodular gravity, i.e.,%
\begin{equation}
\Delta _{\mu \nu }^{\circ }\mathfrak{A}=\Psi _{\mu \nu }^{\circ }\mathfrak{A}%
\text{ \ \ \ and \ \ \ }\square \mathfrak{A}=\Psi \mathfrak{A}.  \label{abc2}
\end{equation}%
with $\Psi _{\mu \nu }^{\circ }=R_{\mu \nu }^{\circ }$. We stress that in
general $\Psi _{\mu \nu }\neq R_{\mu \nu }$. Object $\Psi _{\mu \nu }$ was
defined in the following way (\ref{abc}), by wanting that its traceless
version coincides with $R_{\mu \nu }^{\circ }$ while its trace gives $\Phi
\neq R$. In general $R_{\mu \nu }=\Psi _{\mu \nu }+q_{\mu \nu }\varphi $,
with $4\varphi \equiv R-\mathfrak{A}^{-1}\square \mathfrak{A}$. If $\varphi
=0$, the inequalities change to equalities.

Let us work in vacuum. If $\tau _{\mu \nu }=0$, the objects to put in (\ref%
{abc2}) are%
\begin{equation}
\Psi _{\mu \nu }^{\circ }=-\mathcal{E}_{\mu \nu }\text{ \ \ and \ \ }\Psi =%
\frac{1}{3}\left[ \frac{2f}{\mathfrak{A}}-R-\frac{R}{\mathfrak{A}}\right] .
\label{pot2}
\end{equation}%
First expression is obtained by comparing (\ref{ttv}), (\ref{pot}) with
definition of $\Psi _{\mu \nu }$, while second is generated by taking (\ref%
{ft}) in the vacuum. In the follows, we will apply the formulation developed
here.

We will consider \textrm{FRW} \textit{ansatz}, i.e.,%
\begin{equation}
q_{\mu \nu }dx^{\mu }dx^{\nu }=-dt^{2}+a(t)^{2}\delta _{ij}dx^{i}dx^{j},%
\text{ \ \ }i,j=1,2,3,  \label{frw}
\end{equation}%
with $a(t)$ being the scale factor. $\delta _{ij}$ is Kronecker delta
component. Traceless gravity can be written as follows%
\begin{equation}
R_{\mu \nu }-\frac{1}{\mathfrak{A}}\Delta _{\mu \nu }\mathfrak{A}=\frac{1}{4}%
q_{\mu \nu }\left( R-\Psi \mathfrak{A}\right) ,
\end{equation}%
where we have used (\ref{pot}), (\ref{abc2}) and right expression of (\ref%
{pot2}). If we consider (\ref{frw}) and, then, if we take into account $00$
and $ii$ components, we deduce%
\begin{equation}
3\frac{\partial _{t}\partial _{t}a}{a}+\frac{\partial _{t}\partial _{t}%
\mathfrak{A}}{\mathfrak{A}}=\frac{1}{4}\left( R-\Psi \mathfrak{A}\right) =2%
\frac{\left( \partial _{t}a\right) ^{2}}{a^{2}}+\frac{\partial _{t}\partial
_{t}a}{a}+\frac{\partial _{t}\mathfrak{A}}{\mathfrak{A}}\frac{\partial _{t}a%
}{a},
\end{equation}%
which yields%
\begin{equation}
\left[ \partial _{t}\partial _{t}-H\partial _{t}+2\left( \partial
_{t}H\right) \right] \mathfrak{A}=0.  \label{ap0}
\end{equation}

On another hand, scalaron equation (right expression of (\ref{abc2}))
provides%
\begin{equation}
\left[ \partial _{t}\partial _{t}+3H\partial _{t}+\Psi \right] \mathfrak{A}%
=0.  \label{ap1}
\end{equation}%
If we combine (\ref{ap0}) with (\ref{ap1}), we derive%
\begin{equation}
2\left[ \partial _{t}-2\left( \partial _{t}\ln \mathfrak{A}\right) \right]
H=\Psi .  \label{ghe}
\end{equation}%
Equation (\ref{ghe}) is our guide to study universe expansion for generic
case. Hubble function which satisfies (\ref{ghe}) is%
\begin{equation}
H(t)=\mathfrak{A}^{2}\left[ C+\frac{1}{2}\int \frac{\Psi }{\mathfrak{A}^{2}}%
dt\right] ,  \label{hf}
\end{equation}%
with $C$ constant.

Of course that $H$, $\mathfrak{A}$ and $\Psi $ of (\ref{hf}) must be
compatible with $R=6(\partial _{t}H+2H^{2})$ obtained due to definition of $%
R $, since $\mathfrak{A}$ is given by (\ref{adef}). If we restrict (\ref{ghe}%
) considering $f(R)=R+\mathfrak{a}R^{n}$ with $n\geq 2$ and, then, by taking
vacuum brane and \textrm{RS} fine-tuning, we have%
\begin{equation}
\left[ \partial _{t}-2\left( n-1\right) \left( \frac{\partial _{t}R}{R}%
\right) \right] H=\frac{1}{6n\mathfrak{a}R^{n-1}}\left[ R+\mathfrak{a}\left(
2-n\right) R^{n}\right] .  \label{ghe2}
\end{equation}%
If we take yet \textrm{SSMS} brane, with $n=2$ and $m_{4}^{2}=1/6\mathfrak{a}
$, and if consider too $R=-4Q=6(\partial _{t}H+2H^{2})$, we will obtain from
(\ref{ghe2}) our modified first Starobinsky-Friedmann equation:%
\begin{equation}
\mathfrak{\digamma }=m_{4}^{2}Q.  \label{bui}
\end{equation}%
If we choice yet Starobinski bulk/brane, (\ref{bui}) yields%
\begin{equation}
\mathfrak{\digamma }=\mathfrak{\digamma }_{y=0}^{bulk},\text{ \ \ with \ \ }%
\mathfrak{\digamma }^{bulk}\equiv \left[ \frac{5}{8\left( 1+2\mathfrak{b}%
\mathcal{R}\right) }\mathcal{O}_{\ast }-\frac{1}{4}\left( \boxminus -\frac{1%
}{16}\mathcal{R}\right) \right] \mathcal{R}.  \label{c2}
\end{equation}%
In general $\mathfrak{\digamma }^{bulk}\mathfrak{\propto }\Theta _{\ast
}-(1/4)\Pi _{\ast }^{\left[ \mathcal{F}\right] }\mathcal{F}$, such that we
can think $\mathfrak{\digamma }$ as remnant from $5D$ scalaron field.

\subsection{Cosmological Solutions}

We have obtained formal aspects from the theory presented. Now, we will
study two possible cosmological solutions associated with it. In the first
case, we will work the $f(R)$ Starobinsky theory. In the second application,
we will investigate the \textrm{NOO} equation which governs the cosmological
evolution to obtain an example with dark radiation correction. We will
choice the Starobinsky scale factor for this and, by applying the method of
the reconstruction, we will obtain which $f(R)$ generates it.

\subsubsection{$f(R)$ Starobinsky theory}

We have seen that the cosmological evolution in Starobinsky model satisfies
the expression (\ref{bui}). Let us rewrite (\ref{bui}) as follows%
\begin{equation}
\mathfrak{\digamma }=-3m_{4}^{2}H^{2}-\frac{3}{2}\iota m_{4}^{2}\partial
_{t}H.  \label{bui2}
\end{equation}%
The object $\iota $ can assume the values $1$ or $0$. By comparing (\ref%
{bui2}) with (\ref{Re}), if $\iota =0$ we recover the Starobinsky theory
without extra dimension while $\iota =1$ provides our correction. In
inflationary epoch we have $\digamma \simeq 18H^{2}\partial _{t}H$, such
that putting it in (\ref{bui2}), we obtain the following solution%
\begin{equation}
H_{\pm }^{\iota }\simeq H_{0}-\frac{m_{4}^{2}}{12}(t-t_{0})\pm \frac{m_{4}}{%
12}\sqrt{m_{4}^{2}(t-t_{0})^{2}+12\iota }.
\end{equation}%
We must note that if $\iota =0$ we obtain the usual Starobinsky inflation
for $H_{-}^{0}$. The scale factor associated with $H_{\pm }^{\mathcal{E}}$\
is given by%
\begin{eqnarray}
a_{\pm }^{\iota } &\simeq &a_{0}\exp \left[ H_{0}(t-t_{0})-\frac{m_{4}^{2}}{%
12}(t-t_{0})^{2}\pm \frac{m_{4}}{24}(t-t_{0})\sqrt{m_{4}^{2}(t-t_{0})^{2}+12%
\iota }\right]  \notag \\
&&\times \exp \left[ \pm \frac{m_{4}\iota }{2}\ln \left( m_{4}\sqrt{%
m_{4}^{2}(t-t_{0})^{2}+12\iota }+m_{4}^{2}(t-t_{0})\right) \right] .
\label{sia}
\end{eqnarray}%
Again, we recover the Starobinsky scale factor if $a_{-}^{0}$. Thus, (\ref%
{sia}) describes the cosmological evolution in the inflation epoch for $f(R)$
Starobinsky theory.

\subsubsection{Starobinsky scale factor: application of the reconstruction
method with dark radiation influence}

We will assume the unimodular constraint for induced metric, i. e., $\sqrt{-q%
}=1$, by taking the line element as follows%
\begin{equation}
q_{\mu \nu }dx^{\mu }dx^{\nu }=-a(\tau )^{-6}d\tau ^{2}+a(\tau )^{2}\delta
_{ij}dx^{i}dx^{j},\text{ \ \ }i,j=1,2,3.
\end{equation}%
In \cite{noo} has been shown that on this metric, the equation which governs
the cosmological evolution is given by%
\begin{equation}
\frac{d^{2}}{d\tau ^{2}}\left( d_{R}f\right) +2h\frac{d}{d\tau }\left(
d_{R}f\right) +\left( 6h^{2}+2\frac{d}{d\tau }h\right) \left( d_{R}f\right) +%
\frac{1}{2a^{6}}\left( \rho +P\right) =0,  \label{nooe}
\end{equation}%
with $h(\tau )=a(\tau )^{-1}\partial _{\tau }a(\tau )$. Since we have a dark
radiation correction given by $\Delta T_{\mu \nu }^{\circ }=-\mathcal{E}%
_{\mu \nu }$, we rewrite $\Delta \rho =-\mathcal{E}_{00}$ and $\Delta P=-%
\mathcal{E}_{ii}$\ in (\ref{nooe}), where we will consider now $\rho +\Delta
\rho $ and $P+\Delta P$. Usually, in the \textrm{SMS} brane/bulk, the dark
radiation can be express as follows \cite{mukohyama2}%
\begin{equation}
\mathcal{E}_{0}{}^{0}=-\frac{q_{\mathcal{E}}}{a^{4}}\text{ \ \ and \ \ }%
\mathcal{E}_{i}{}^{i}=\frac{q_{\mathcal{E}}}{3a^{4}},  \label{tc}
\end{equation}%
with $q_{\mathcal{E}}$ the \textit{dark radiation tidal} charge. Let us
consider (\ref{tc}) generalized to%
\begin{equation}
\mathcal{E}_{0}{}^{0}=-\frac{q_{\mathcal{E}}}{a^{n}}\text{ \ \ and \ \ }%
\mathcal{E}_{i}{}^{i}=\frac{q_{\mathcal{E}}}{3a^{n}}.
\end{equation}%
Therefore, (\ref{nooe}) can be rewritten as%
\begin{equation}
\frac{d^{2}}{d\tau ^{2}}\left( d_{R}f\right) +2h\frac{d}{d\tau }\left(
d_{R}f\right) +\left( 6h^{2}+2\frac{d}{d\tau }h\right) \left( d_{R}f\right) +%
\frac{1}{2a^{6}}\left( \rho +P\right) =\frac{q_{\mathcal{E}}}{2}\left( \frac{%
1}{3a^{n+4}}+\frac{1}{a^{n+12}}\right) .  \label{nooe2}
\end{equation}

By following the Ref. \cite{noo}, we will choice%
\begin{equation}
H(t)=H_{0}-\frac{m_{4}^{2}}{6}\left( t-t_{0}\right)
:a(t)=a_{0}e^{H_{0}\left( t-t_{0}\right) -m_{4}^{2}\left( t-t_{0}\right)
^{2}/12}\overset{t,t_{0}\ll H_{0}^{-1}}{\simeq }a_{0}e^{H_{0}\left(
t-t_{0}\right) +m_{4}^{2}tt_{0}/6},  \label{it}
\end{equation}%
to obtain the $f(R)$ gravity which generates (\ref{it}). The quantities $%
m_{4}$, $t_{0}$\ and $H_{0}$\ are arbitrary constants. On the conformal time 
$\tau $, where $d\tau =a(t)^{3}dt$, the scale factor and $h(\tau )$ become%
\begin{equation}
a(\tau )=\left( \frac{\mathcal{N}}{2}\right) ^{1/3}\left( \tau -\tau
_{0}\right) ^{1/3}\text{ \ \ and \ \ }h(\tau )=\frac{1}{3\left( \tau -\tau
_{0}\right) },  \label{csf}
\end{equation}%
where we have used $\mathcal{N}=6H_{0}+m^{2}t_{0}$. As have been shown in 
\cite{noo}, the values $H_{0}=0,00000016$, $m_{4}=0,053$, $t_{0}=10^{-3}$sec
and $a_{0}=1$ lead for the spectral index $n_{s}\simeq 0,966$, which is
compatible with the Planck result. By using (\ref{csf}) in (\ref{nooe2}) and
considering $\rho =0=P$, we obtain%
\begin{equation}
\frac{d^{2}}{d\tau ^{2}}\left( d_{R}f\right) +\frac{2}{3\left( \tau -\tau
_{0}\right) }\frac{d}{d\tau }\left( d_{R}f\right) =\frac{q_{\mathcal{E}}}{6}%
\left( \frac{2}{\mathcal{N}}\right) ^{\frac{n+4}{3}}\left[ \frac{1}{\left(
\tau -\tau _{0}\right) ^{\frac{n+4}{3}}}+\frac{3\cdot 2^{8/3}}{\mathcal{N}%
^{8/3}\left( \tau -\tau _{0}\right) ^{\frac{n+12}{3}}}\right] .  \label{bcd}
\end{equation}%
The equation (\ref{bcd}) provides the following solution%
\begin{equation}
d_{R}f=1+3\left( \tau -\tau _{0}\right) ^{1/3}+\frac{3q_{\mathcal{E}}}{2}%
\left( \frac{2}{\mathcal{N}}\right) ^{\frac{n+4}{3}}\left[ \frac{\left( \tau
-\tau _{0}\right) ^{\frac{2-n}{3}}}{n^{2}-3n+2}+\frac{6\cdot 2^{8/3}\left(
\tau -\tau _{0}\right) ^{\frac{-n-6}{3}}}{\mathcal{N}^{8/3}\left(
n^{2}+13n+22\right) }\right] .  \label{bcd2}
\end{equation}%
We have taken the integration constants equal to one. The relationship
between the conformal time and the curvature scalar is given by \cite{noo}%
\begin{equation}
\tau -\tau _{0}=\frac{2\cdot 3^{3/2}}{\mathcal{N}^{4}}R^{3/2}.  \label{ctc}
\end{equation}%
By putting (\ref{ctc}) in (\ref{bcd2}), we obtain%
\begin{equation}
d_{R}f=1+\frac{2^{1/3}\cdot 3^{3/2}}{\mathcal{N}^{4/3}}R^{1/2}+\frac{6\cdot
3^{\left( 2-n\right) /2}q_{\mathcal{E}}}{\mathcal{N}^{4-n}}\left[ \frac{%
R^{\left( 2-n\right) /2}}{n^{2}-3n+2}+\frac{2\mathcal{N}^{8}R^{-\left(
n+6\right) /2}}{27\left( n^{2}+13n+22\right) }\right] .  \label{pu}
\end{equation}%
Finally, if we integrate (\ref{pu}) in $R$, we can find the $f\left(
R\right) $ theory which generates the scale factor (\ref{csf}). Explicitly,
we find%
\begin{equation}
f\left( R\right) =R+\frac{3^{5/6}\cdot 2^{2/3}}{\mathcal{N}_{0}^{4/3}}%
R^{3/2}+q_{\mathcal{E}}\widetilde{f}\left( R\right) ,
\end{equation}%
with $\widetilde{f}\left( R\right) $ given by%
\begin{equation}
\widetilde{f}\left( R\right) =\frac{36}{3^{^{n/2}}\mathcal{N}^{4-n}}\left[ 
\frac{R^{-\left( n-4\right) /2}}{\left( 4-n\right) \left( n^{2}-3n+2\right) }%
-\frac{2\mathcal{N}^{8}R^{-\left( n+4\right) /2}}{27\left( n+4\right) \left(
n^{2}+13n+22\right) }\right] .  \label{etc}
\end{equation}%
The \textrm{NOO} case is obtained simply by taking $q_{\mathcal{E}}=0$.

Let us analyze specific situations of (\ref{etc}) for $n\geq 0$. For
instance, if $n=0$ we have the theories $\sim R^{2}$ and $\sim -R^{-2}$. For 
$n=1,2,4$, the first term in (\ref{etc}) diverges. Actually, if $n=4$ we
obtain%
\begin{equation}
f\left( R\right) =R+\frac{3^{5/6}\cdot 2^{2/3}}{\mathcal{N}_{0}^{4/3}}%
R^{3/2}+\frac{q_{\mathcal{E}}}{3}\left[ \ln R-\frac{\mathcal{N}^{8}}{810}%
R^{-4}\right] ,
\end{equation}%
by taking $n=4$ in (\ref{pu}). We must note that for $n\geq 5$, the first
term becomes negative. Thus, if $q_{\mathcal{E}}>0$, our correction provides
a negative contribution for \textrm{NOO} theory and, if $q_{\mathcal{E}}<0$%
,\ this contribution becomes positive. In the follows, we provide a table
detailing some possible scenarios for $0\leq n\leq 9$:%
\begin{equation*}
\begin{tabular}{|l|l|l|l|}
\hline
$n$ & $\widetilde{f}\left( R\right) $ & $n$ & $\widetilde{f}\left( R\right) $
\\ \hline
$0$ & $\frac{36}{\mathcal{N}^{4}}\left( \frac{1}{8}R^{2}-\frac{2\mathcal{N}%
^{8}}{2376}R^{-2}\right) $ & $6$ & $-\frac{36\mathcal{N}^{2}}{3^{^{3}}}%
\left( \frac{1}{40}R^{-1}+\frac{2\mathcal{N}^{8}}{36720}R^{-5}\right) $ \\ 
\hline
$3$ & $\frac{36}{3^{^{3/2}}\mathcal{N}}\left( \frac{1}{2}R^{1/2}-\frac{2%
\mathcal{N}^{8}}{13230}R^{-7/2}\right) $ & $7$ & $-\frac{36\mathcal{N}^{3}}{%
3^{^{7/2}}}\left( \frac{1}{90}R^{-3/2}+\frac{2\mathcal{N}^{8}}{48114}%
R^{-11/2}\right) $ \\ \hline
$4$ & $\frac{1}{3}\ln R-\frac{\mathcal{N}^{8}}{2430}R^{-4}$ & $8$ & $-\frac{%
36\mathcal{N}^{4}}{3^{^{4}}}\left( \frac{1}{168}R^{-2}+\frac{2\mathcal{N}^{8}%
}{61560}R^{-6}\right) $ \\ \hline
$5$ & $-\frac{36\mathcal{N}}{3^{^{5/2}}}\left( \frac{1}{12}R^{-1/2}+\frac{2%
\mathcal{N}^{8}}{29646}R^{-9/2}\right) $ & $9$ & $-\frac{36\mathcal{N}^{5}}{%
3^{^{9/2}}}\left( \frac{1}{280}R^{-5/2}+\frac{2\mathcal{N}^{8}}{77220}%
R^{-13/2}\right) $ \\ \hline
\end{tabular}%
\end{equation*}

\section{Conclusion}

We have developed two different aspects associated with effective $f(R)$
branewords. The first has been general rules to create it from $\mathcal{F(R)%
}$-bulk. By wanting to generate a \textquotedblleft copy\textquotedblright\
of the bulk dynamics, we have developed the \textrm{CDC} concept, in which
geometrical reduction does not affect dynamical structure of the left side,
such that $\mathcal{F(R)}$ in $5D$ is fully analog to $f(R)$ in $4D$.
Similar idea to generate effective $f(R)$-brane was developed in \cite%
{aguilar}, with projection stipulated by applying \textrm{CGC} directly in $%
\mathcal{F}$: $\mathcal{F(R)}=f(R+\mathcal{K})$. Thus, we have obtained a
new dynamical projection class: $\mathcal{F(R)}\Rightarrow f(R)$. As future
study, must be found a physical principle which provides the correct
projection, as done in \cite{no} to solve brane predictability problem.

In the second step, we have taken a specific case to obtain $f(R)$%
-unimodular gravity. We showed that the derived equations are formally
identical with obtained in \cite{noo}. Two applications were obtained and a
full study about these solutions will be examined in a future paper.

\end{document}